\documentclass[aps, prl, reprint, superscriptaddress]{revtex4-1}
\usepackage[utf8]{inputenc}
\usepackage{amsmath}
\usepackage{amsfonts}
\usepackage{amssymb}
\usepackage{graphicx}
\usepackage{lineno}
\begin{document}
\title{Interaction of Strain and Nuclear Spins in Silicon: Quadrupole Effects on Ionized Donors}

\author{David P. Franke}
\email{david.franke@wsi.tum.de}
\affiliation{Walter Schottky Institut and Physik-Department, Technische Universität München, Am Coulombwall 4, 
85748 Garching, Germany}
\author{Florian M. Hrubesch}
\affiliation{Walter Schottky Institut and Physik-Department, Technische Universität München, Am Coulombwall 4, 
85748 Garching, Germany}
\author{Markus K\"unzl}
\affiliation{Walter Schottky Institut and Physik-Department, Technische Universität München, Am Coulombwall 4, 
	85748 Garching, Germany}
\author{Hans-Werner Becker}
\affiliation{RUBION, Ruhr-Universität Bochum, Universitätsstraße 150, 44780 Bochum, Germany}
\author{Kohei M. Itoh}
\affiliation{School of Fundamental Science and Technology, Keio University, 3-14-1 Hiyoshi, 
	Kohoku-ku, Yokohama 223-8522, Japan}
\author{Martin Stutzmann}
\affiliation{Walter Schottky Institut and Physik-Department, Technische Universität München, Am Coulombwall 4, 
	85748 Garching, Germany}
\author{Felix Hoehne}
\affiliation{Walter Schottky Institut and Physik-Department, Technische Universität München, Am Coulombwall 4, 
	85748 Garching, Germany}
\author{Lukas Dreher}
\affiliation{Walter Schottky Institut and Physik-Department, Technische Universität München, Am Coulombwall 4, 
85748 Garching, Germany}
\author{Martin S. Brandt}
\affiliation{Walter Schottky Institut and Physik-Department, Technische Universität München, Am Coulombwall 4, 
	85748 Garching, Germany}

\begin{abstract}
	The nuclear spins of ionized donors silicon have become an interesting quantum resource due to their very long coherence times. Their perfect isolation, however, comes at a price, since the absence of the donor electron makes the nuclear spin difficult to control. We demonstrate that the quadrupolar interaction allows to effectively tune the nuclear magnetic resonance of ionized arsenic donors in silicon via strain and determine the two nonzero elements of the S-tensor linking strain and electric field gradients in this material to $S_{11}=1.5\times 10^{22}$ V/m$^2$ and $S_{44}=6\times 10^{22}$ V/m$^2$. We find a stronger benefit of dynamical decoupling on the coherence properties of transitions subject to first-order quadrupole shifts than on those subject to only second-order shifts and discuss applications of quadrupole physics including mechanical driving of magnetic resonance, cooling of mechanical resonators and strain-mediated spin coupling.
\end{abstract}

\maketitle
The potential of computing based on quantum mechanics has triggered a quest for a scalable quantum bit or qubit technology with local control and long coherence times. Recent studies on the nuclear spin of ionized phosphorus donors in silicon have demonstrated ultra-long coherence times $T_2$ exceeding tens of minutes even at room temperature \cite{steger_quantum_2012, saeedi_room-temperature_2013}. Together with the advantages of modern silicon semiconductor technology, these times make such nuclear spins a prime candidate for the implementation of a quantum memory \cite{boehme_nuclear-spin_2012}. To enable storage of quantum information on longer timescales, the implementation of error correction is vital \cite{shor_scheme_1995}, requiring local control of single qubits, as well as two- or many-qubit interactions \cite{dennis_topological_2002}. Such a local control can be achieved by the combination of the global use of nuclear magnetic resonance (NMR) with a local tuning by gates \cite{kane_silicon-based_1998}, which has proven challenging even for the neutral phosphorus donor system \cite{wolfowicz_conditional_2014}. 
For ionized donors, the absence of the corresponding donor electron impedes any electronic control which would otherwise be mediated by the so-called hyperfine interaction of electron and nuclear spins, leading to a near perfect isolation of the ionized donor nuclear spin. 
In contrast to phosphorus, however, the heavier group V donors with a nuclear spin $I>1/2$ such as arsenic possess a nonzero quadrupole moment which interacts with electric field gradients \cite{wasylishen_nmr_2012}. Here, we show that this additional interaction can be manipulated by elastic strain applied to the host crystal and can very successfully be used to mechanically tune the properties of ionized donor nuclear spins. This paves the way for scalable local addressing of these qubits, as strain can be applied locally by piezo-actuators on the scale of nanometers \cite{alexe_self-assembled_2006}. Similarly, mechanical resonators in the required MHz frequency range are routinely fabricated in silicon in the framework of microelectromechanical systems (MEMS) and should be able to realize spin-spin couplings mediated by phonons, ultimately generating two- or many-qubit gates \cite{yeo_strain-mediated_2014}. Furthermore, while the lack of a stable silicon isotope with $I>1/2$ has prohibited accoustic nuclear resonance studies into quadrupole interaction in this material, the experiments reported here finally allow to measure the effectiveness with which electric field gradients are generated on substitutional lattice sites in silicon under the application of strain.

The initialization of ionized donor qubits can be realized by processes of spin-selective ionization based on the hyperfine interaction of electron and nuclear spins in the neutral charge state, which in most cases allows for an electrical read out of the qubit at the same time \cite{dreher_nuclear_2012, pla_high-fidelity_2013, hoehne_submillisecond_2015}. 
In our experiments, we detect the resonance frequencies of the nuclear spins of ionized arsenic donors As$^{+}$ making use of the Pauli blockade formed in weakly coupled spin pairs. These pairs form between the neutral donors As$^0$ with an electron spin $S=1/2$ and the paramagnetic oxygen-vacancy complex (SL1) in monocrystalline silicon exhibiting $S=1$ \cite{franke_spin-dependent_2014-1, franke_spin-dependent_2014} [Fig.~1 (a)]. Because of the Pauli exclusion principle, the electron transfer from the donor level to the energetically lower SL1 level strongly depends on the spin symmetry of the two partners (parallel and antiparallel). This leads to an effective polarization into the long-lived parallel configuration in the steady state. Changing the spin symmetry by manipulating the spin state of either partner promotes the charge transfer, which is detected as a resonant increase in the conductivity, attributed to an Auger-recombination-type process \cite{franke_spin-dependent_2014-1}.

The samples used in this work are silicon with natural isotope composition which was implanted with As$^+$ ions, resulting in a maximum As concentration of about $7\times 10^{19}$ cm$^{-3}$ about 70 nm below the surface. The samples were not annealed in order to preserve the defect centers created upon implantation  \cite{franke_spin-dependent_2014}, which leads to a low dopant activation ratio as confirmed by the observation of hyperfine-resolved ESR peaks. The samples were contacted with an interdigit structure, biased with typically 8 V and illuminated with a red (wavelength 635 nm) light-emitting diode. All measurements were performed at 8 K and at a fixed microwave frequency (9.74 GHz) in a dielectric resonator for pulsed ENDOR. After the application of mw pulses, a resonant transient increase of the conductivity of the sample is recorded with a fast digitizer card.

\begin{figure}
	\centering
	\includegraphics[width=\linewidth]{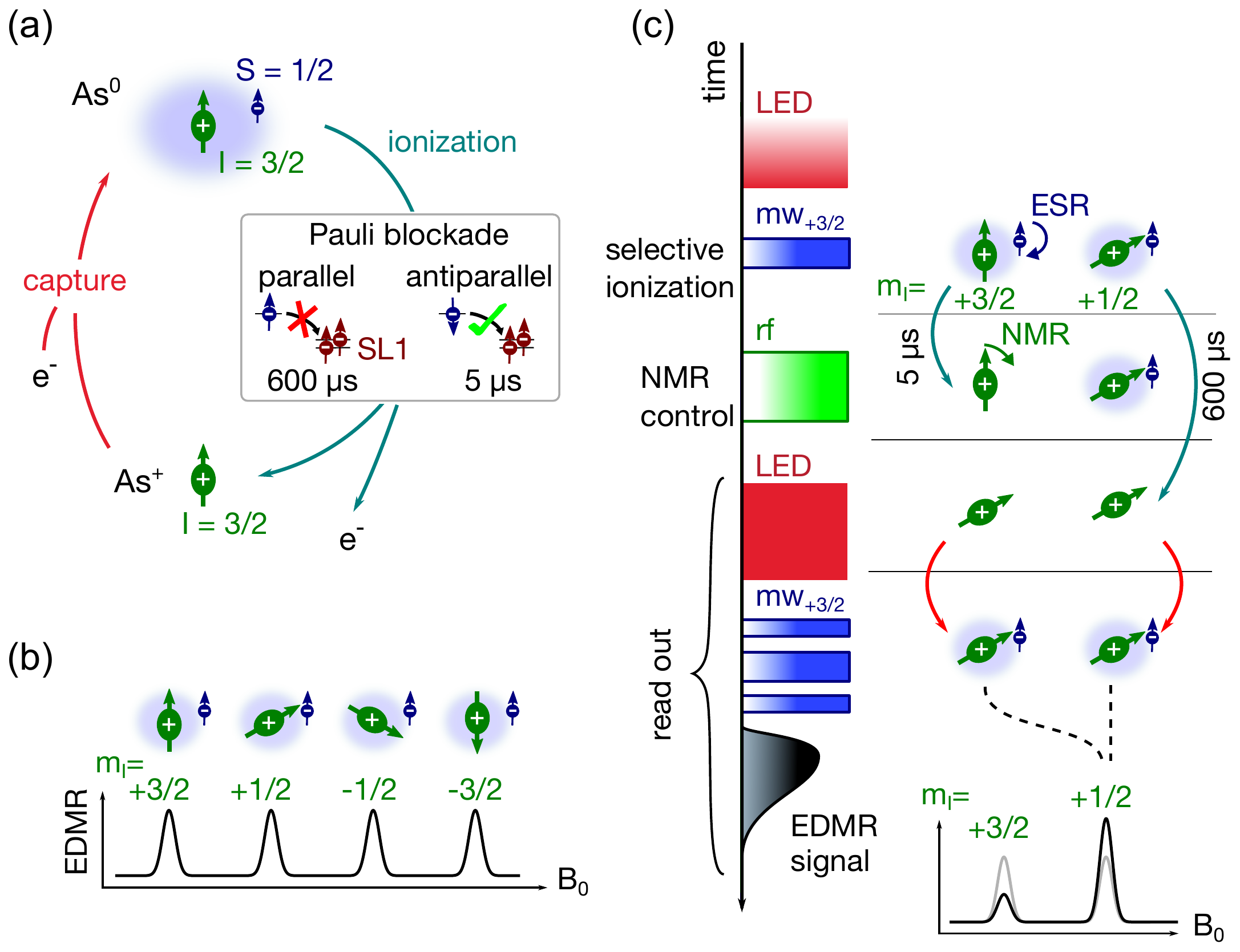}
	\caption{
		Manipulation and electrical detection of $^{75}$As nuclear spins based on spin-dependent recombination.
		(a) Charge states As$^0$ and As$^+$ of an arsenic donor with nuclear and electron spin states denoted by green and blue arrows, respectively. The ionization is governed by a Pauli blockade of the donor electron spin with a paramagnetic defect (SL1).
		(b) Schematic spectrum showing the enhanced conductivity caused by electrons ejected to the conduction band during the ionization process \cite{franke_spin-dependent_2014-1}.
		(c) Evolution of the charge and spin states of the $^{75}$As donors during the ED ENDOR sequence. See text for details.}
	\label{fig:fig1}
\end{figure}

The electron spin resonance (ESR) of As$^0$ splits into four different levels, in first order separated by the hyperfine interaction of $198.35$ MHz \cite{feher_electron_1959}. Each ESR line corresponds to one of the four nuclear spin eigenstates $m_I=-3/2\dots3/2$ of the $I=3/2$ $^{75}$As nucleus [Fig.~1 (b)], allowing the selective ionization and read out of the nuclear spin state of the ensemble. In combination with radio frequency (rf) pulses, this allows us to perform electrically detected electron nuclear double resonance (ED ENDOR) \cite{hoehne_electrical_2011, malissa_room-temperature_2014}, as shown schematically in Fig.~\ref{fig:fig1} (c). At the beginning of the pulse sequence, a selective microwave (mw) pulse lifts the Pauli blockade and thus promotes the ionization of As$^0$ in a particular nuclear spin state, here $m_I=+3/2$ (\emph{selective ionization}). After the fast antiparallel recombination (on a timescale of 5 $\mu$s \cite{franke_spin-dependent_2014}), the ionized donor spins are highly polarized and can be manipulated by rf pulses (\emph{NMR control}), changing the nuclear spin state. This leads to a nuclear polarization after the slower parallel recombination (600 $\mu$s) has ionized the remaining donors. After the whole ensemble is transferred to the neutral charge state via illumination by a light-emitting diode (LED) and electron capture, the polarization is measured by application of an electron spin echo sequence and detection of the ensuing current transient through the sample (electrically detected magnetic resonance, EDMR)(\emph{read out}) \cite{boehme_theory_2003, hoehne_lock-detection_2012, hoehne_time_2013}. When ionization and detection are performed on the same resonance ($m_I=3/2$), the observed EDMR signal is reduced in case of successful NMR manipulation.

\begin{figure}
	\centering
	\includegraphics[width=\linewidth]{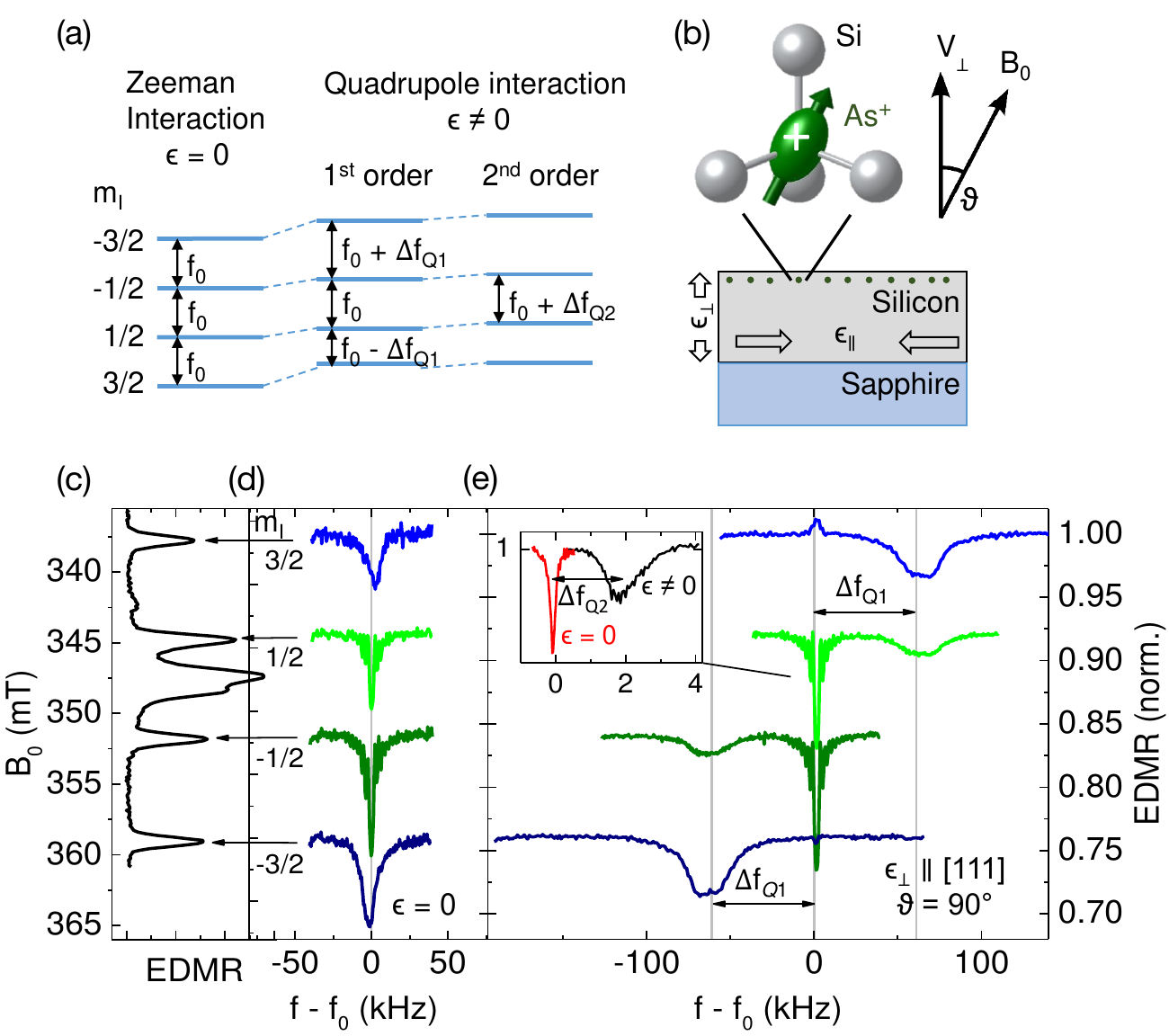}
	\caption{Demonstration of the strain-induced tuning of ionized donors As$^+$.
		(a) Effects of the first- and second-order quadrupole interaction on the magnetic sublevels of As$^+$.
		(b) Generation of strain via the differences in the thermal expansion coefficients of Si and a sapphire substrate. As donors are generated by shallow implantation.
		(c) EDMR spectrum of As$^0$ in Si. 
		(d) NMR transitions of As$^+$ in unstrained Si after ionization on each of the four As$^0$ EDMR peaks.
		(e) Corresponding NMR transitions for  As$^+$ in strained Si showing first-order quadrupole shifts. The inset shows a high-resolution spectrum revealing the second-order quadrupole shift of the $m_I=1/2 \leftrightarrow -1/2$ transition.}
	\label{fig:fig2}
\end{figure}

The spin states of an ionized arsenic donor As$^+$ in silicon are subject to the nuclear Zeeman interaction, which splits the energy levels in an external magnetic field $B_0$ (assumed to be oriented in the $z$ direction), and the interaction between the quadrupole moment $q$ and an electric field gradient $V_{ij}=\partial E_i/\partial j$, where $E_i$ are the electric field components and $i,j$ are the coordinates $x,y,z$. 
The total Hamiltonian describing the nuclear spin components $I_x$, $I_y$, and $I_z$ can be written as
\[
\mathcal{H}_\mathrm{n}=-hf_0 I_z+\frac{eq}{12}\sum\limits_{i,j}V_{ij}[(I_iI_j+I_jI_i)-\delta_{ij}\frac{5}{2}]
\mathrm{ ,}
\]
where $f_0=\mu_\mathrm{n}g_\mathrm{n}B_0/h$ is the Larmor frequency of the nuclear spin, $\mu_\mathrm{n}$ and $g_\mathrm{n}$ are the nuclear magneton and the nuclear $g$-factor, resp., $h$ is Planck's constant and $e$ is the elementary charge \cite{wasylishen_nmr_2012}. For a cubic crystal symmetry as in unstrained Si, the electric field gradients cancel out and the spin states are separated by the Zeeman interaction only. In this case, the frequencies of the three allowed NMR transitions ($\Delta m_I=\pm1$) are degenerate [Fig.~\ref{fig:fig2} (a)]. 

To measure the NMR frequencies of As$^+$, we record ED ENDOR spectra at four magnetic fields, corresponding to the ESR lines of As$^0$ [Fig.~2 (c)], and thereby detect the relative occupancy of the four nuclear spin states of As$^+$. When monitoring these occupancies while sweeping the frequency of the rf pulse, a reduction is expected on resonance with the NMR transitions. Therefore, after ionization of, e.g., the $m_I=3/2$ state, the signature of the transition $m_I=3/2\leftrightarrow 1/2$ is anticipated, while for ionization on $m_I= 1/2$ the measurement is sensitive to two transitions, $m_I=3/2\leftrightarrow 1/2$ and $m_I=1/2\leftrightarrow -1/2$.
Experimentally, in unstrained samples only one peak is observed in each of the four NMR spectra [Fig.~2 (d)]. This is in agreement with the expected degeneracy of the three NMR transitions for zero quadrupole interaction [Fig.~2 (a)]. From the observed resonance positions $g_\mathrm{n}=0.9558(2)$ can be extracted, corresponding to a chemical shift of $-0.40(2)\%$ with respect to the value for free As nuclei \cite{stone_table_2005}. 

A uniaxial strain $\epsilon_\perp$ applied to the cubic Si crystal reduces the cubic symmetry and results in an electric field gradient $V_\perp$ which interacts with the arsenic nuclear spin, giving a nonzero quadrupole energy $hf_Q=V_{\perp}eq$. The change in resonance frequency is described by (i) a first order shift $\Delta f_{Q1}\propto f_Q$ of the ``outer" transitions $m_I=3/2\leftrightarrow 1/2$ and $-3/2\leftrightarrow -1/2$, and (2) a second order shift $\Delta f_{Q2}\propto f_Q^2/f_0$ of all four resonance lines, including and most easily observed on the ``inner" transition $m_I = 1/2\leftrightarrow -1/2$ [Fig.~2 (a)].
In our proof-of-principle experiments, strain is applied by cementing a thin silicon sample to a sapphire substrate which results in uniaxial strain $\epsilon_{\perp}$ at low temperature, caused by the different thermal expansion coefficients of the materials [Fig.~2 (b)]. A precise in-situ measurement of the resulting strain is provided by the change of the hyperfine interaction of electron and nuclear spin of the neutral donor in the case of strain applied in the $[100]$ direction of the crystal \cite{wilson_electron_1961}. We find an expansion of $\epsilon_{\perp}\sim 2.9\times 10^{-4}$. This corresponds to an in-plane compression of $\epsilon_{\parallel} \sim -3.8\times 10^{-4}$, in agreement with the thermal expansion coefficients for Si and c-plane sapphire \cite{ibach_thermal_1969, lucht_precise_2003}. Due to the high reproducibility of the strain observed in all of the $(100)$ Si-sapphire stacks investigated and the isotropy of the thermal expansion in Si, we assume that the applied strain is equal in $(111)$ stacks, to which this method to measure $\epsilon_{\perp}$ cannot be applied because of the symmetry of the conduction band minima \cite{wilson_electron_1961}.
\begin{figure}
	\centering
	\includegraphics[width=\linewidth]{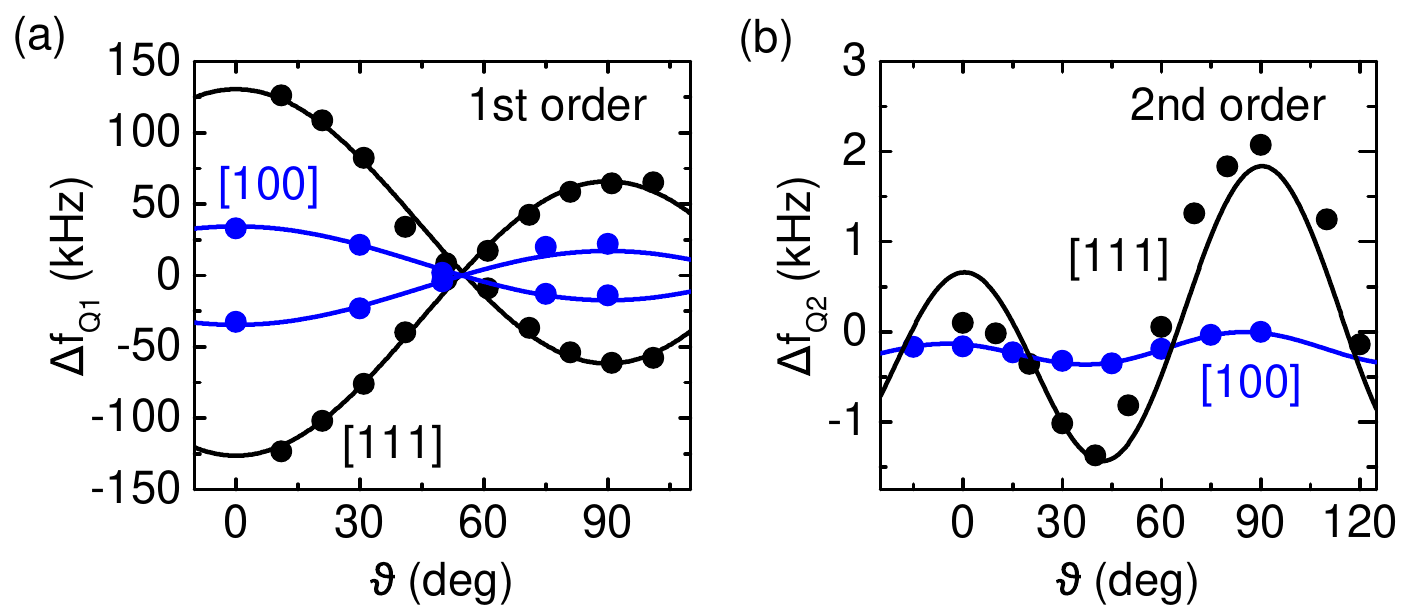}
	\caption{Angular dependences of the first-order (a) and second-order (b) quadrupole shifts on the relative orientation of the magnetic field $B_0$ and the electric field gradient $V_{\perp}$ generated in $[111]$ (black) and $[100]$ (blue) direction.}
	\label{fig:fig3}
\end{figure}
Figure 2 (e) shows that for $(111)$ stacks, the first order shift is clearly observed, separating the peaks corresponding to inner and outer transitions. The outer transition resonance peaks are shifted by well above one linewidth, while the sharper peaks of the inner transition are barely moved with respect to $f_0$ on the frequency scale of Fig.~2 (e). The lineshapes of the sharp lines close to $f_0$ in Fig.~2 (d) and (e) are dominated by a pattern reflecting the distribution of frequencies given by the Fourier transform of the $400~\mu$s NMR square pulse used, while the shapes of the resonances observed for $m_I=\pm3/2$ are asymmetrically broadened. For the nominally unstrained samples, this is most likely due to strain induced by the contact structure \cite{kawamura_strain-induced_2010}, the additional asymmetry of the broad lines in Fig.~2 (e) is attributed to an inhomogeneity of the strain generation. The weak resonances at $f-f_0=0$ in the top and bottom spectra most likely originate from a dynamic crosstalk, where changes in the polarization of the $m_I=1/2$ and $-1/2$ states influence the polarization of the monitored states via relaxation.

When using longer rf pulses, a higher spectral resolution can be reached. For a pulse length of $6$ ms, the second order shift to the inner transition becomes evident, as shown in the inset of Fig.~\ref{fig:fig2} (e), where the resonance line is shown with (black trace) and without strain (red trace) in direct comparison. Since the linewidth of this transition is smaller by more than two orders of magnitude compared to the outer transitions, the resonance is shifted by more than one linewidth as well, even though the shift $\Delta f_{Q2}$ is significantly smaller in absolute numbers. Note that the quadratic dependence of $\Delta f_{Q2}$ on $f_Q$ will reduce the negative influence of small fluctuations for this shift. To further confirm our interpretation of the observed shifts of the resonances as resulting from quadrupole interactions, the characteristic dependence of both first and second order shifts on the angle $\vartheta$ between strain axis and magnetic field was measured [Fig.~3 (a) and (b)]. When treated as a perturbation of the Zeeman interaction, the expected quadrupolar shifts for $I=3/2$ are given by \cite{wasylishen_nmr_2012}
\begin{align*}
	\Delta f_{Q1}&=\frac{f_Q}{2}\left[\frac{1}{2}(3\cos^2\vartheta-1)\right]\text{ and}\\
	\Delta f_{Q2}&=\frac{3f_Q^2}{64f_0}(1-\cos^2\vartheta)(9\cos^2\vartheta-1)\text{ ,}
\end{align*}
which fits the observed angular dependence very well [solid lines in Fig.~3 (a) and (b)].

The quadrupole shift is caused by an anisotropic electric field gradient which interacts with the arsenic nuclear spin and is proportional to the applied strain, $V_i=S_{ij}\cdot \epsilon_j$ (using Voigt notation). The tensor $S_{ij}$ describing the proportionality in cubic crystals has two nontrivial components, $S_{11}$ and $S_{44}$, which quantify the effect of strain (along the principle axes) and shear, respectively \cite{nye_physical_1985}. They can be investigated separately by measuring samples strained in [100] and [111] crystal directions. When fitting the angular dependence functions to the experimental data for first and second order, an $f_Q$ of $76(5)$ kHz and $255(3)$ kHz can be extracted from the shifts in [100] and [111] samples, respectively. Hence, $S_{11}=1.5(3)\times 10^{22}~\mathrm{V/m}^2$ and  $S_{44}=6.8(20)\times 10^{22}~\mathrm{ V/m}^2$ using $q=31.4$ fm$^2$ for $^{75}$As \cite{stone_table_2005}.

While there are no previous measurements of $S$ in Si due to the absence of stable isotopes with nonzero quadrupole interaction, we note that the obtained values are similar to those reported from acoustic nuclear resonance on $^{73}$Ge in Ge \cite{sundfors_nuclear_1979}, both with respect to the absolute values of $S_{11}$ and $S_{44}$ and to their relative size.
This correspondence is expected due to the similar crystal structure of Si and Ge and the equal core electron configuration of As and Ge. Experiments on Sb and Bi will show how universal the $S$-parameters obtained here are for silicon.

\begin{figure}
	\centering
	\includegraphics[width=\linewidth]{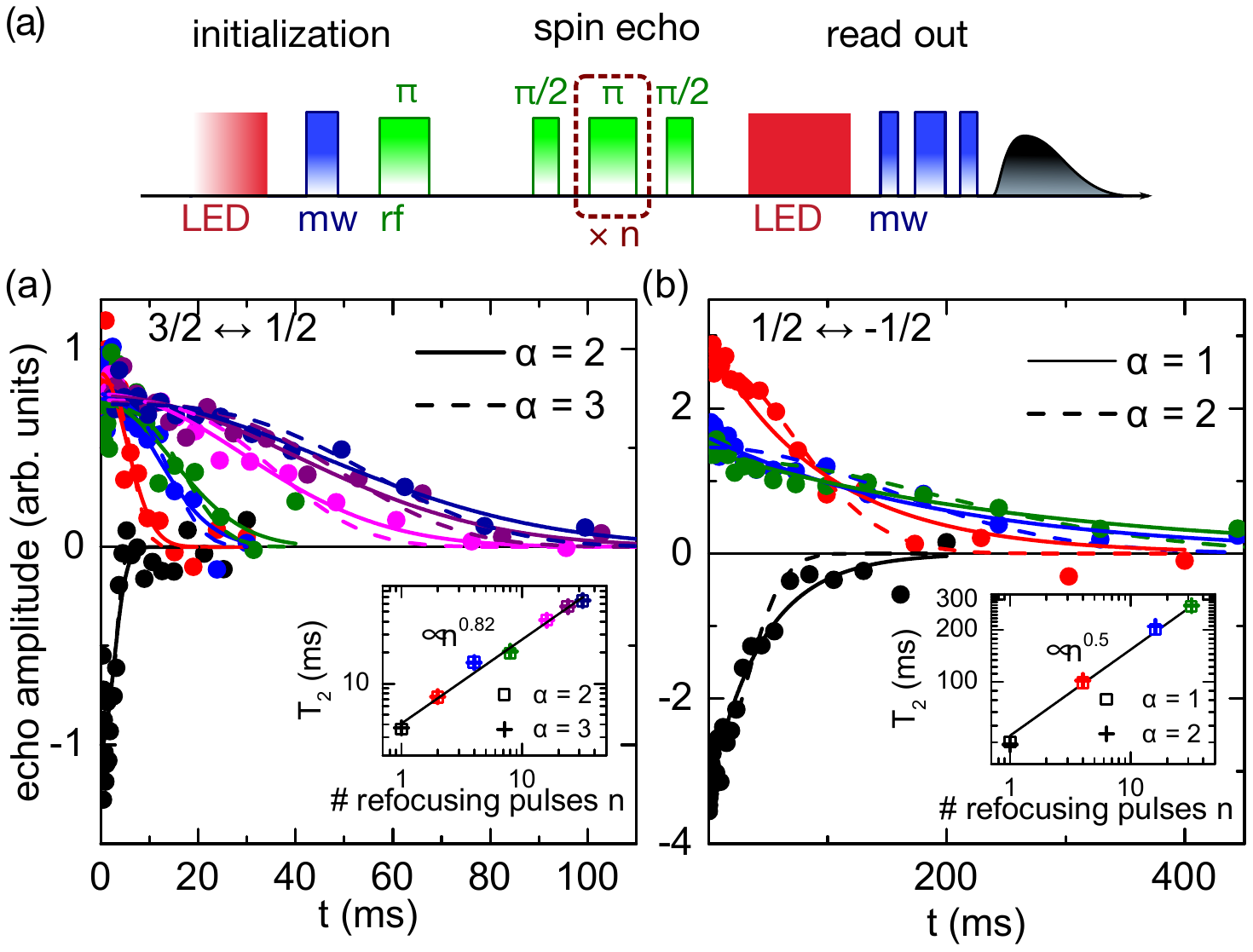}
	\caption{(a) Pulse sequence used for coherence time measurements. Nuclear spin echo decays of the $3/2\leftrightarrow 1/2$ (b) and $1/2\leftrightarrow -1/2$ (c) subsystems of the As$^+$ nuclear spins. Colors denote different numbers of refocusing pulses $n$ in the dynamical decoupling sequence as indicated in the insets, dashed and solid lines correspond to fits with superexponential decays exp[$-(t/T_2)^{\alpha}$] with different exponents $\alpha$. The insets show the extracted decay constants $T_2$ as a function of $n$.}
		\label{fig:fig4}
	\end{figure}


To study the influence of the quadrupole interaction on the decoherence of the nuclear spins, we measure the coherence time $T_2$ using a spin echo sequence with an additional $\pi/2$ projection pulse which is inserted after the rf $\pi$-pulse [Fig.~4 (a)]. Due to the larger coupling strength of the first order effects, we expect a bigger influence on the coherence time observed on the outer transitions. The decay of the echo amplitude [black traces in Fig.~4 (b) and (c)] confirms the expected difference, giving $T_2\sim 44(5)$ ms for the inner and $T_2\sim 3.7(3)$ ms for the outer transition. These values are found to be mostly unchanged after the application of stress in our experiments. While $T_2$ for the inner transition is comparable to the coherence time of ionized phosphorus donors in silicon with natural isotope composition \cite{dreher_nuclear_2012}, the coherence time of the outer transition is reduced, in accordance with the expected stronger coupling to the phonon bath. By application of a Carr-Purcell dynamical decoupling sequence \cite{carr_effects_1954} with 32 refocusing pulses, $T_2$ is extended to $T_2=275(10)$ ms and $T_2=64(5)$ ms for the inner and outer transitions, respectively. While the enhancement with the number of applied decoupling pulses $n$ scales $\propto n^{0.5}$ for the inner transition [inset of Fig.~4 (b)], as would be expected for $1/f$ noise \cite{medford_scaling_2012}, a different scaling $\propto n^{0.8}$ is observed for the outer transition [inset of Fig.~4 (c)], indicating that a different process dominates the decoherence in this case. This could be connected to, e.g., phonon number fluctuations \cite{bennett_phonon-induced_2013} or field gradients introduced by charged defects. However, the stronger scaling with $n$ suggests that, for a large number of pulses, this additional decoherence mechanism can be made negligible. Even longer coherence times could most likely be reached using isotopically purified silicon to overcome the interaction with $^{29}$Si nuclear spins in the host crystal \cite{itoh_isotope_2014}.

Coming back to the application of the strain interaction to the tuning of qubits, 
with strains of $5\times 10^{-5}$ which can be realized by electrically addressable piezo-actuators \cite{dreher_electroelastic_2011}, shifts of the $3/2\leftrightarrow 1/2$ transition of the order of 18 kHz could be realizable. Compared to the corresponding transition in our proof-of-principle samples in Fig.~2 (e), this strain corresponds to a shift by one full linewidth. Going to smaller ensembles or single As$^+$ spins, strain inhomogeneities will be reduced or removed, allowing shifts of many linewidths and an excellent selectivity, even without the use of isotopical engineering. 
Due to the strain connected to the oscillation of, e.g., a nano-mechanical beam, the quadrupole interaction also renders possible the coupling of the nuclear spin to micro-mechanical modes. An estimate based on elasticity theory for a doubly clamped beam \cite{bennett_phonon-induced_2013} with dimensions $1\times 0.1\times 0.1~\mu$m$^3$ results in a coupling constant $g/2\pi\approx 18$ Hz, which is of the same order of magnitude as the observed coherence time. Other possible application of strain interactions include the implementation of mechanical driving of nuclear spin resonance in Si and the nuclear spin-based cooling of mechanical oscillators, allowing to achieve higher displacement and force sensitivity \cite{mamin_sub-attonewton_2001, kepesidis_phonon_2013}, as well as the generation of squeezed spin states \cite{kitagawa_squeezed_1993}.

\begin{acknowledgments}
The authors would like to thank Manabu Otsuka for the sample characterization, and Pierre-André Mortemousque and \L ukasz Cyrwi\'nski for fruitful discussions. This work was supported financially by DFG via SFB 631 and SPP 1601.
\end{acknowledgments}

\bibliography{bib3}
\end{document}